\documentclass[aps,prl,twocolumn, groupedaddress,showpacs]{revtex4}
\usepackage[centertags]{amsmath}
\usepackage{graphicx,amssymb}

\begin{document}
\title{Possible origin of the reduced  ordered moment in iron
  pnictides: a Dynamical Mean Field Theory study}

\author{Hunpyo Lee}
\author{Yu-Zhong Zhang}
\author{Harald O. Jeschke}
\author{Roser Valent\'\i}
\affiliation{Institut f\"ur Theoretische Physik, Goethe-Universit\"at
Frankfurt, Max-von-Laue-Stra{\ss}e 1, 60438 Frankfurt am Main, Germany}
\date{\today}

\begin{abstract}
  We investigate the phase diagram of a two-band frustrated Hubbard
  model in the framework of dynamical mean field theory.  While a
  first-order phase transition occurs from a paramagnetic (PM) metal
  to an antiferromagnetic (AF) insulator when both bands are equally
  frustrated, an intermediate AF metallic phase appears in each band
  at different $U_c$ values if only one of the two bands is
  frustrated, resulting in continuous orbital-selective phase
  transitions from PM metal to AF metal and AF metal to AF
  insulator, regardless of the strength of the Ising Hund's coupling. We argue that 
  our minimal model  calculations capture the frustration behavior in the
  undoped iron-pnictide superconductors as well as local quantum fluctuation
  effects and that the
 intermediate phases  observed in our results are possibly
  related to the puzzling AF metallic state with small staggered magnetization
  observed in these systems as well as to the
  pseudogap features observed in optical experiments.
\end{abstract}

\maketitle

The subtle interplay among magnetism, superconductivity, multiorbital
effects and structure is a major subject of debate in the recently
discovered iron pnictide superconductors~\cite{Kamihara08}.  While,
similar to high-$T_{\text{c}}$ cuprate superconductors, magnetically
mediated pairing was proposed to dominate the superconducting
state~\cite{Mazin08PRL}, the nature of magnetism in undoped iron
pnictides is still unclear~\cite{Mazin08PRB}.  The experimentally
observed iron  ordered 
 moment in the antiferromagnetic phase is too small, compared to that obtained
from density functional theory (DFT) calculations.  Various DFT
studies have shown that this value is strongly dependent on the
details of the calculations and on the lattice
structure~\cite{Opahle08,Mazin08PRB}.  Very recently, a local density approximation (LDA)+U
calculation explained the small magnetic moment in terms of large
magnetic multipoles without analyzing the nature of the phase
transition~\cite{Cricchio09}.  A few alternative proposals are based
on a localized picture where a frustrated one-band Heisenberg model is
considered~\cite{Yildirim08,Si08,Han09,Schmidt09}.  However, the
multi-band and itinerant nature of iron pnictides are overlooked in
such approaches.  Furthermore, existing dynamical mean field theory (DMFT) and
LDA + DMFT studies~\cite{Haule08,Craco08,Haule09,Skornyakov09,Ishida10} for iron pnictides
were performed in the paramagnetic state and did not consider the magnetic ordering. Therefore, a proper microscopic theory for the magnetism in iron pnictides is
still missing.

Analysis of recent Fe~$3d$ transfer integrals obtained from
downfolding the bandstructure of a few iron-based
superconductors~\cite{Miyake09} always shows the existence of  weakly
frustrated (like $d_{xy}$) and  highly frustrated (like
$d_{yz}$/$d_{zx}$) orbitals for all cases due to the hopping mediated
by a pnictogen or chalcogen ion.  Such behavior suggests that a
minimal model for exploring the role of frustration on the
magnetism of the iron pnictides should
be a two-band model with one unfrustrated and one frustrated band.
The question to be posed is whether an AF metallic state with small ordered
magnetic moment can emerge out of the interplay between frustrated and
unfrustrated bands.

In order to investigate this issue, we consider in the present work a
two-band half-filled Hubbard model with different degrees of band
frustration.  We will demonstrate that while the AF metallic state is
absent when both bands are equally frustrated, an AF metallic state
with small magnetization is present when the frustration in one of the
bands is turned off. Moreover, we identify a pseudogap region and show
that it originates from the small AF moment which is due to the
interplay between frustrated and unfrustrated bands.

The Hamiltonian we study is
\begin{equation}\begin{split}
  H &=-\sum_{\langle ij\rangle m\sigma}  t_m c^{\dagger}_{im\sigma}
  c_{jm\sigma} -\sum_{\langle ij'\rangle m\sigma}  t_m^{\prime} c^{\dagger}_{im\sigma}
  c_{j'm\sigma} \\
    &+\,U\sum_{im}n_{im\uparrow} n_{im\downarrow} +\sum_{i\sigma\sigma'}\big(U'-\delta_{\sigma \sigma'} J_z\big)n_{i1\sigma}
  n_{i2\sigma'}\,,
\label{eq:hamiltonian}
\end{split}\end{equation}
where $c_{im\sigma}(c^{\dag}_{im\sigma})$ is the annihilation
(creation) operator of an electron with spin $\sigma$ at site $i$ and
band $m$. $t_m$ ($t_m^{\prime}$) is the hopping matrix element between
site $i$ and nearest-neighbor (NN) site $j$ (next nearest-neighbor
(NNN) site $j'$). $t_m^{\prime}=0$ for the unfrustrated band. For
simplification, we neglect inter-band hybridizations.  $U$ and $U'$
are, respectively, intra-band and inter-band Coulomb interaction
integrals and $J_z n_{i1 \sigma}n_{i2 \sigma}$ is the Ising-type
Hund's coupling term. In our calculations we set $U'=\frac{U}{2}$ and
$J_z=\frac{U}{4}$ and ignore the spin-flip and pair-hopping processes.
For the solution of this model we employ DMFT~\cite{Georges96}
 which includes the local quantum fluctuation
effects and we perform the calculations on the Bethe lattice. 
  The DMFT self-consistency equations with inclusion of the
N\'eel state are given as~\cite{Zitzler04,Georges96}
\begin{eqnarray}
 G_{0,A,\sigma}^{-1}=i\omega_n + \mu - t_m^2 G_{B,\sigma} - t_m^{\prime 2}G_{A,\sigma}, \\
 G_{0,B,\sigma}^{-1}=i\omega_n + \mu - t_m^2 G_{A,\sigma} - t_m^{\prime 2}G_{B,\sigma},
\end{eqnarray}
where $\mu$ is the chemical potential, $\omega_n$ is the Matsubara
frequency and magnetizations of $A$ and $B$ sublattices are in
opposite directions. As impurity solver, a weak-coupling
continuous-time quantum Monte Carlo algorithm was
employed~\cite{Rubtsov05,Lee08}.

We first consider the two-band Hubbard model with magnetic frustration
in both bands at half-filling. Previous DMFT calculations done on the
frustrated one-band Hubbard model~\cite{Zitzler04} with frustration
strength $t'/t=0.58$, showed the existence of a first-order phase
transition from paramagnetic (PM) metal to AF insulator. For
comparison with this one-band case, we set in our two-band model
$t_m=1$ and $t'_m=0.58$ for $m=1,2$.  The bandwidth $W=4.624$ is
determined as $W = 4\sqrt{t^2 + {t'}^2}$.

\begin{figure}
\includegraphics[width=0.4\textwidth]{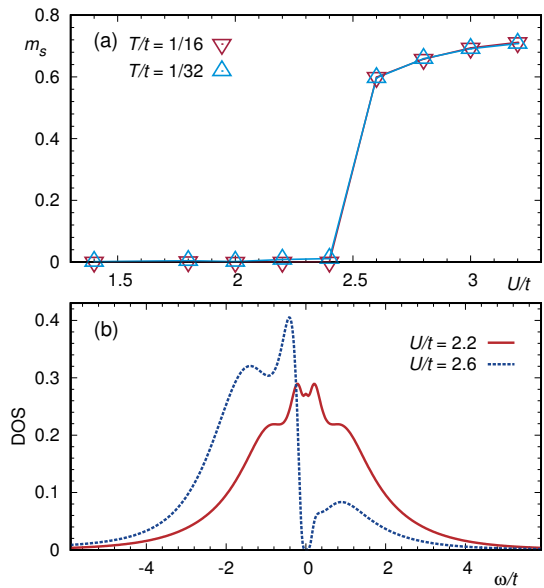}
\caption{(a) Staggered magnetization $m_s$ as a function of $U/t$
for
  $T/t=1/16$ and $1/32$. A first-order PM to AF transition,
  recognizable from a jump of $m_s$, is present at the critical
  interaction strength $U_c/t=2.4$. (b) Density of states (DOS) for
  the spin up species on an A lattice site as a function of frequency
  for $U/t=2.2$ and $2.6$ and $T/t=1/32$.} \label{fig:FF_MAG_DOS}
\end{figure}

In Fig.~\ref{fig:FF_MAG_DOS} we present the results for the staggered
magnetization $m_s$ as a function of $U/t$ for $T/t=1/16$ and
$T/t=1/32$ ( Fig.~\ref{fig:FF_MAG_DOS}~(a)) and the density of states
(DOS) at $U/t=2.2$ and $U/t=2.6$ for $T/t=1/32$ (
Fig.~\ref{fig:FF_MAG_DOS}~(b)).
Below $U/t=2.4$ in Fig.~\ref{fig:FF_MAG_DOS}~(a) the staggered
magnetization for both temperatures is negligibly small, indicating a
PM state. As the interaction $U/t$ is increased, for both temperatures
a jump is detected around the critical value of $U_c/t=2.4$ and the
system goes into an AF state. The discontinuous behavior suggests a
first-order phase transition. These results are very similar to those
obtained for the one-band Hubbard model with
frustration~\cite{Zitzler04}. In order to analyze the metal-insulator
transition, we present in Fig.~\ref{fig:FF_MAG_DOS}~(b) the DOS close
to the critical $U_c/t$ where we employed the maximum entropy method
for analytic continuation. In the PM state at $U/t=2.2$, the observed
finite DOS at the Fermi level ($\omega=0$) indicates a metal. In the
AF state at $U/t=2.6$, the spin-up and spin-down DOS on the same
sublattice become unequal and the spin-up (spin-down) DOS on
sublattice A and the spin-down (spin-up) DOS on B are pairwise equal
due to the development of the AF moments. Due to the Coulomb
interaction strength $U/t$ and the appearance of AF ordering, the
system shows insulating behavior with opening of a small gap at the
Fermi level ($\omega=0$). Comparing the results of the magnetically
frustrated one-band Hubbard model with the two-band model where an
orbital degree of freedom is involved, we find that the phase diagrams
of both models are qualitatively the same. Such a model cannot
reproduce the magnetic behavior of the Fe pnictides and also should
not be relevant for the phase diagram of the Mott insulator
V$_2$O$_3$~\cite{Zitzler04,Imada98}.

Now let us consider the two-band system in which frustration is turned
off for one of the bands.  This model should mimic the observed
behavior in downfolding calculations~\cite{Miyake09} for the Fe
pnictides. We set $t_1=1$ and $t_1'=0$ for the unfrustrated band and
$t_2=1$ and $t_2'=0.65$ for the frustrated one.  The bandwidths for
unfrustrated and frustrated bands are $W_1=4.0$ and $W_2=4.77$,
respectively.
\begin{figure}
\includegraphics[width=0.4\textwidth]{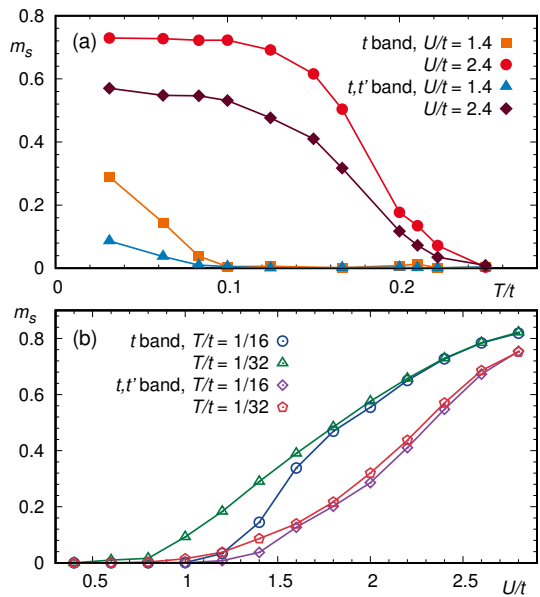}
\caption{(a) Staggered magnetization $m_s$ for the unfrustrated band
  ($t$ band) and the frustrated band ($t,t'$ band) (a) as a function
  of $T/t$ for $U/t=1.4$ and $2.4$ and (b) as a function of $U/t$ for
  $T/t=1/16$ and $1/32$.  A continuous transition with a smooth
  increase of $m_s$ is observed as a function of $U/t$.}
\label{fig:T_U_dependence}
\end{figure}
In Fig.~\ref{fig:T_U_dependence}~(a) we show the behavior of the
staggered magnetization $m_s$ as a function of temperature $T/t$ for
two different interaction strengths $U/t$. At $U/t=1.4$ ($U/t=2.4$),
the staggered magnetization $m_s>0$ for both frustrated and
unfrustrated bands is detected as temperature decreases below the
N\'eel temperature around $T_N/t\simeq0.1$ ($T_N/t\simeq0.22$) where
the system undergoes a PM-AF phase transition. The staggered
magnetization increases more rapidly in the unfrustrated band than in
the frustrated one. In Fig.~\ref{fig:T_U_dependence}~(b) we show the
staggered magnetization as a function of interaction strength $U/t$
for two temperature values. We find a smooth increase of the
magnetization with $U/t$ for both bands and for both temperatures.
Unlike the case of magnetic frustration in both bands where a
first-order phase transition was observed
(Fig.~\ref{fig:FF_MAG_DOS}~(a)), this smoothly increasing behavior of
the staggered magnetization $m_s$ is a strong evidence of the
existence of continuous phase transitions. It also suggests the
existence of an AF metal where the small staggered magnetization is
not sufficient for opening a full gap; this is what we investigate
next.
\begin{figure}
\includegraphics[angle=-90,width=0.4\textwidth]{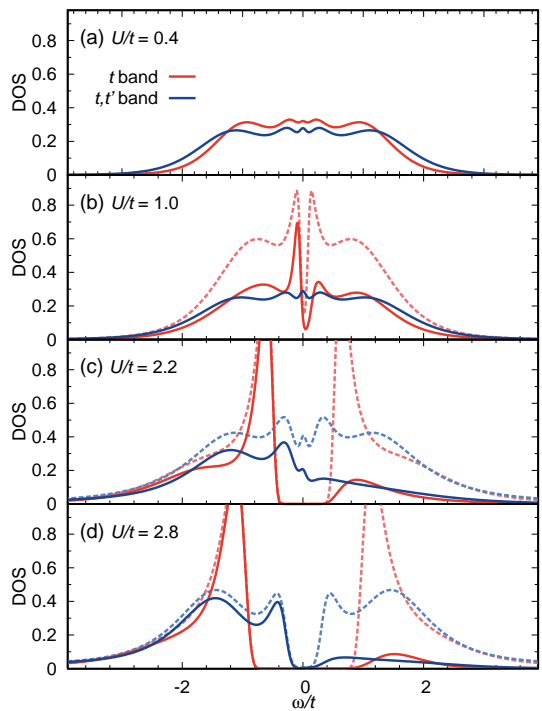}
\caption{Spin-up DOS on sublattice A (solid lines) as a function of
  frequency analyzed at $T/t = 1/32$ for (a) $U/t=0.4$, (b) $U/t=1.0$,
  (c) $U/t=2.2$ and (d) $U/t=2.8$. Also shown (dotted lines) is the
  total spin-up DOS when AF order occurs.} \label{fig:DOS}
\end{figure}

To analyze the metal-to-insulator transition, we present in
Fig.~\ref{fig:DOS} the spin-up DOS on the A site for four
representative values of $U/t$ at a fixed temperature of $T/t=1/32$.
The PM metallic state in both bands is observed at $U/t=0.4$ (see
Fig.~\ref{fig:DOS}~(a)). As the interaction is increased to $U/t=1.0$
(see Fig.~\ref{fig:DOS}~(b)), a magnetic transition occurs where the
frustrated band ($t,t'$ band) remains in a PM state while an AF
metallic state is present in the unfrustrated band ($t$ band). At this
value of $U/t$, the small moment only opens a pseudogap.  When we
increase the interaction, an orbital selective metal-to-insulator
transition occurs, and at $U/t=2.2$ (see Fig.~\ref{fig:DOS}~(c)), an
AF metal in the frustrated band coexists with an AF insulator in the
unfrustrated band. Finally, in the strong-coupling region at $U/t=2.8$
(see Fig.~\ref{fig:DOS}~(d)), both bands are in AF insulating states.

\begin{figure}
\includegraphics[width=0.4\textwidth]{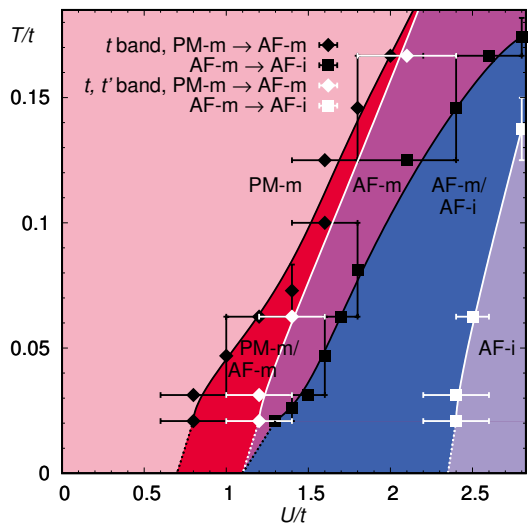}
\caption{Magnetic phase diagram for the two-band Hubbard model where
  an unfrustrated band ($t$ band, $t=1$,$t'=0$) and a frustrated band
  ($(t,t')$ band, $t=1$, $t'=0.65$) coexist. The phase boundaries'
  error bars are also shown.}
\label{fig:phasediagram}
\end{figure}

In Fig.~\ref{fig:phasediagram} we plot the phase diagram $T/t$ versus
$U/t$ for the Hamiltonian~\eqref{eq:hamiltonian}. The PM metal, AF
metal and AF insulator phases are present in both bands, but the
critical values $U_c/t$ of the unfrustrated band are smaller than
those of the frustrated one. The N\'eel temperature increases as a
function of $U/t$. The intermediate AF metals show pseudogap behavior
in the DOS as a precursor of gap opening (see {\it e.g.}
Fig.~\ref{fig:DOS}~(b) for the unfrustrated band) due to the
continuous phase transitions induced by a continuous change of
magnetization (see Fig.~\ref{fig:T_U_dependence}).  This is in
contrast to a first-order Mott transition dominated by strong
correlations where an abrupt gap opening is observed. The pseudogap
features obtained here could account for the experimentally observed
optical conductivity behavior of the new Fe-based
superconductors~\cite{Qazilbash09}.

\begin{figure}
\includegraphics[width=0.4\textwidth]{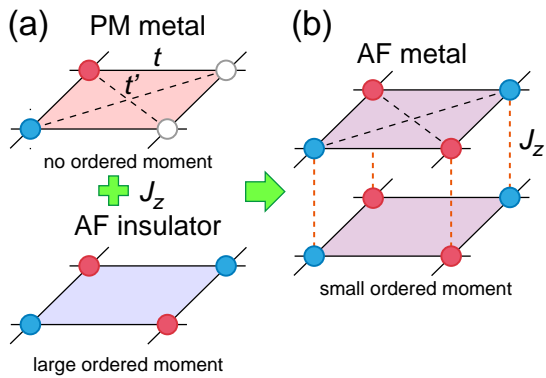}
\caption{Schematic picture for the mechanism of the appearance of an
  AF metal with small ordered moment from the coupling of a frustrated
  band (PM metal) with an unfrustrated band (AF insulator). The idea
  is sketched on a square lattice with regard to iron-based
  superconductors. Red (dark gray) and blue (light gray) sites suggest
  spin configurations, with white sites representing frustration.}
\label{fig:cartoon}
\end{figure}

We visualize the mechanism of the appearance of an AF metallic phase
with small antiferromagnetic ordered  moment in Fig.~\ref{fig:cartoon}.
Without coupling between frustrated and unfrustrated bands, the ground
state of the unfrustrated band on a square lattice shows AF insulating
behavior with high magnetic ordered moment as soon as the interaction $U>0$
due to perfect nesting while that of the frustrated band exhibits
nonmagnetic metallic behavior below a critical interaction $U_c/t$ as
frustration prevents perfect nesting (see
Fig.~\ref{fig:cartoon}~(a)). As the Hund's rule coupling ($J_z$),
which favors ferromagnetic arrangement of spins on the same site, is
switched on between these two bands, the spins in the nonmagnetic
frustrated band tend to order antiferromagnetically as in the
unfrustrated band, and the itinerant electrons have the tendency to
localize since hopping between nearest-neighbor sites violates Hund's
rule.  On the other hand, the spins in the AF unfrustrated band are
affected by frustration due to the Hund's coupling to the frustrated
band.  Therefore, they follow the spin arrangement in the frustrated
itinerant band and become more delocalized (see
Fig.~\ref{fig:cartoon}~(b)).  Such an interplay between frustrated and
unfrustrated bands results in a reduction of the antiferromagnetic
ordered  moments and
therefore of the gap amplitude in the density of states, explaining
the additional AF metallic phases (see Fig.~\ref{fig:phasediagram}).

In summary, we have studied the frustrated two-band Hubbard model at
half-filling and have shown that a first-order phase transition
separating a PM metal from an AF insulator occurs if both bands are
equally frustrated.  On the other hand, by considering one band
frustrated and turning off frustration in the second band, orbital
selective continuous phase transitions occur in both bands first from
a PM metal to an AF metal and then from an AF metal to an AF
insulator. This leads to new phases where either both bands are AF
metals, or the frustrated band is an AF metal while the unfrustrated
one is still a PM metal, or the frustrated band is already AF
insulating while the unfrustrated one is still an AF metal.  These new
phases may be directly relevant for the magnetism of the new
iron-based superconductors where the small ordered magnetic moments observed
in the stripe-type antiferromagnetic phase may result from an interplay between frustrated and unfrustrated
bands.  Furthermore, the pseudogap behavior in the AF metal state is
closely related to the optical conductivity features of iron-based
superconductors~\cite{Qazilbash09}.

The new phases involving AF metallic states appear in a wide range
of interaction parameters, indicating that our model can be applied
to a large family of iron-based superconductors with different
interaction strengths. In the present work we showed the case of one
unfrustrated band coupled with one frustrated band with
$t'_2/t_2=0.65$ but we have checked a few more cases at $T/t=1/16$
by tuning to stronger frustrations in the frustrated band
($t'_2/t_2=0.8$) or even by changing the unfrustrated band to be
weakly frustrated ($t'_1/t_1=0.2$). In both cases we find solutions
of AF metals, underlining the relevance of the investigated model
for the new iron-based superconductors. Furthermore, we have checked
that AF metallic states also exist at both $J_z=U/8$ and $J_z=7U/24$
with the constraint of $U=U'+2J_z$ in addition to the value of
$J_z=U/4$ we present in this work.  Our model calculations show that it is the
coupling of strongly frustrated with weakly frustrated bands which induces a
reduced antiferromagnetic ordered moment, and this should be applicable to
many (more than two) bands with different degrees of frustration as is the
case in the iron pnictides.

While we believe that our model calculations qualitatively capture the
central physics of AF metal with small ordered magnetic moment observed
experimentally in undoped iron-based superconductors as well as the
nature of the phase transitions, further investigations have to be
done by including all five Fe $3d$ orbitals with realistic inter-band,
intra-band hybridizations and various fillings on the frustrated
square lattice in order to allow for quantitative comparisons between
experiments and theory.

Our results on the model with two equally frustrated bands also show that this model
is insufficient for explaining the physics of V$_2$O$_3$ contrary to
previous suggestions~\cite{Zitzler04}, and inclusion of other degrees
of freedom like phonons may be necessary.

{\it Acknowledgments}.- We thank I. Opahle for useful discussions and
the Deutsche Forschungsgemeinschaft for financial support through the
SFB/TRR~49 and Emmy Noether programs.

\end{document}